\documentclass[a4paper,11pt,nohyper]{JHEP3}
\usepackage{amsmath,latexsym,amssymb}
\usepackage{graphicx}
\usepackage[latin1]{inputenc}
\usepackage{subfigure}
\usepackage{nicefrac}
\usepackage{braket}
\usepackage{comment}
\usepackage{definitions}

\usepackage{framed}

\title{\boldmath
Axionic symmetry gaugings in ${\cal N}=4$ supergravities and their
higher-dimensional origin \unboldmath}

\author{Jean-Pierre~Derendinger${}^{\clubsuit}$,
P.~Marios~Petropoulos${}^{\spadesuit\; \clubsuit}$ and Nikolaos~Prezas${}^{\diamondsuit}$\\

  \begin{itemize}
  
\item  Institut de Physique, Universit\'e de Neuch\^atel\\
Breguet 1, 2000 Neuch\^atel, Switzerland
    
  \item   Centre de Physique Théorique, Ecole Polytechnique, CNRS\footnote{Unité
      mixte UMR 7644.} \\
    91128 Palaiseau, France
     
\item CERN PH-TH\\
1211 Genève, Switzerland
  \end{itemize}

\bigskip

E-mail: \email{jean-pierre.derendinger@unine.ch,
marios@cpht.polytechnique.fr, nikolaos.prezas@cern.ch}}

\abstract{We study the class of four-dimensional ${\cal N}=4$
supergravities obtained by gauging the axionic shift and axionic
rescaling symmetries. We formulate these theories using the
machinery of embedding tensors, characterize the full gauge
algebras and discuss several specific features of this family of
gauged supergravities. We exhibit in particular a generalized
duality between massive vectors and massive two-forms in four
dimensions, inherited from the gauging of the shift symmetry. We
show that these theories can be deduced from higher dimensions by
a Scherk--Schwarz reduction, where a twist with respect to a
non-compact symmetry is required. The four-dimensional generalized
duality plays a crucial role in identifying the higher-dimensional
ascendent.}

 \preprint{NEIP--07--01\\
  CPHT-RR015.0307\\
 CERN-PH-TH 2007/057\\}

\begin{document}

\setcounter{footnote}{0}
\renewcommand{\thefootnote}{\arabic{footnote}}
\setcounter{section}{0}
\section{Introduction}

The effective field theories describing the low-energy dynamics of
typical string and M-theory compactifications are plagued with
massless  scalar fields that hamper any attempt to contact
four-dimensional phenomenology. One way of eliminating some of
these unwanted massless scalars is the introduction of fluxes in
the internal compactification space (see \cite{Grana:2005jc} for a
review). From the effective field theory perspective, turning on
fluxes corresponds to a ``gauging" of the original theory obtained
without fluxes. The term gauging refers to the fact that a
subgroup of the global duality symmetry of the original theory is
promoted to a local gauge symmetry. Simultaneously, part of the
original Abelian gauge symmetry is promoted to a non-Abelian one 
and various fields, including the scalars, acquire minimal
couplings to the gauge fields. The connection with the moduli
stabilization issue stems from the fact that for extended
supergravity theories, the only way to generate a potential is
through the gauging. The resulting theories are known as gauged
supergravities (see \cite{Trigiante:2007ki} for a concise review).

One can envisage a bottom-up approach to the problem of moduli
stabilization where instead of looking for a specific
higher-dimensional background whose corresponding low-energy
effective theory is phenomenologically viable, one first
constructs a gauged supergravity theory with the required
phenomenological properties and then attempts to engineer it from
higher dimensions. Such a programme was initiated in
\cite{Derendinger:2004jn} and  models were proposed there
exhibiting full moduli stabilization.

Evidently such a programme depends crucially on having a good
picture of the ``landscape" of possible gauged supergravity
theories. Therefore, a problem of paramount importance is to
classify and describe all possible gaugings of supergravity in
diverse spacetime dimensions and with various amounts of
supersymmetry. This problem was tackled in a series of
publications \cite{deWit:2002vt, deWit:2003hq,
deWit:2003hr,deWit:2004nw,deWit:2005ub} where the general method
of {\it embedding tensors} was developed.

Ultimately, however, one would like to know that the effective
theory constructed  in the bottom-up approach is consistent,
i.e.~can be embedded in a certain string or M-theory setup.
Although for many classes of gauged supergravities their
higher-dimensional realization is known, for the generic gauged
supergravity there is no recipe -- not even guaranteed existence
of a higher-dimensional origin.
Presumably such an endeavor would require a better understanding
of the classes of string backgrounds dubbed non-geometric, whose
significance has recently been investigated in the framework of
flux compactifications and supergravity theories.

Here, our aim is to contribute in this direction by analyzing a
family of four-dimensional ${\cal N}=4$ gauged supergravities and
explain how they can be obtained from string theory or
higher-dimensional supergravity theories. Gauged supergravity with
${\cal N}=4$ supersymmetry was the arena of
\cite{Derendinger:2004jn} since string theories with 16
supercharges are the starting point of a variety of realistic
string constructions. Recently the most general theory of this
type was explicitly constructed in  \cite{Schon:2006kz} using the
formalism of embedding tensors. We will use this formalism to
study in detail the specific class of ${\cal N}=4$ theories
obtained by gauging the axionic symmetries, namely the
\emph{axionic shifts and rescalings}. Consistency requires that
several $SO(6,6)$ directions are also gauged. The general
structure of the gauge algebra is systematically worked out and it exhibits
the following characteristic property: it is \emph{non-flat} contrary to what
happens in more conventional gaugings.

In order to uncover the higher-dimensional origin of these
gaugings one needs to perform a Scherk--Schwarz reduction of
heterotic supergravity (or of the common sector in general) on a
torus. The crucial ingredient here is a \emph{twist} by a
non-compact duality symmetry of ten-dimensional supergravity.

The identification of the theory obtained from the reduction with
the gauged supergravity under consideration is not at all
straightforward. It relies heavily on a duality between massive
vectors and massive two-forms in four dimensions. This duality is
a necessary extension of the more standard duality between a
massless two-form and an axion scalar field in four dimensions :
it incorporates the Stückelberg-like terms that are generated by
the axionic gauging.

The organization of this paper is as follows. Section
\ref{sec:gengaug} is devoted to a reminder of gauged
\cite{Freedman:1978ra, Gates:1982ct, Gates:1983ha}  ${\cal N}=4$
supergravity in $D=4$ \cite{Das:1977uy, Cremmer:1977tc, Cremmer:1977tt}.
We emphasize the approach of the embedding
tensor following
Refs.~\cite{deWit:2005ub,Schon:2006kz,Derendinger:2006jb}. In
Sec.~\ref{sec:elecgaug} we specialize to the so-called electric
gaugings and we analyze a particular class of algebras for which
we elaborate on the corresponding gauged supergravity. These
theories admit two equivalent formulations related by a duality
between massive vectors and massive two-forms in four dimensions.
In Sec.~\ref{sec:hetered} we move on to higher dimensions. Our aim is
to analyze the ten-dimensional origin of the four-dimensional
theory constructed in Sec.~\ref{sec:elecgaug}. We show that it can
be obtained using a generalized Scherk--Schwarz reduction of
heterotic $\mathcal{N}=1$ $D=10$ supergravity. Furthermore, we
comment on the higher-dimensional origin of other classes of
${\cal N}=4$ gaugings. In Sec.~\ref{sec:conc} we present  our
conclusions and discuss some open problems.

\section{\boldmath Reminder on ${\cal N}=4$ gauged supergravities in $D=4$ \unboldmath}\label{sec:gengaug}

In this section we review ${\cal N}=4$ gauged supergravity following
Ref.~\cite{Schon:2006kz}. After some general remarks on the possible
gauge algebras and the constraints on the gauging parameters
we present for reference the most general bosonic Lagrangian.

\subsection{Gauge algebras and the embedding tensor}\label{sec:gengaugemb}
Four-dimensional ${\cal N}=4$ supergravity has a very restricted
structure. It contains generically the gravity multiplet and
$n$ vector multiplets. The bosonic sector of the theory consists of
the graviton, $n+6$ vectors and $2+6n$ scalars. In the ungauged
version, the gauge group is Abelian, $U(1)^{6+n}$, and there is
no potential for the scalars.

Interactions are induced upon elimination of the auxiliary fields. They
affect the scalars whose non-linearities are captured by a
universal coset manifold \cite{Chamseddine:1980cp, Derendinger:1984zz}
\begin{equation}\label{eq:scaman}
  \mathcal{M}=\frac{SL(2,\mathbb{R})}{U(1)}\times \frac{SO(6,n)}{SO(6)\times
  SO(n)}.
\end{equation}
All fields  are non-minimally coupled to
the Abelian vectors. The gauge kinetic terms have
scalar-field-dependent coefficients, whereas the action is at most
quadratic in the gauge-field strengths with no explicit dependence
on the gauge potentials. For this reason, the
$SL(2,\mathbb{R})\times SO(6,n) \subset Sp(12+2n,\mathbb{R})$
symmetry of the scalar manifold is globally realized as a U-duality
symmetry. Although the scalar manifold survives any deformation of
the plain theory triggered by gaugings, the U-duality is broken as
a consequence of the introduction of non-Abelian field strengths
and minimal couplings, all of which depend explicitly on the gauge
potentials.

The duality group acts as a symmetry of the field equations and
the Bianchi identities of the gauge fields. In the standard
formulation of supergravity only a subgroup of it is realized
off-shell as a genuine symmetry of the Lagrangian. This includes
the $SO(6,n)$ plus a two-dimensional non-semi-simple subalgebra of
$SL(2,\mathbb{R})$ generated by axionic shifts and axionic rescalings. The
third transformation in $SL(2,\mathbb{R})$, corresponding to the
truly electric-magnetic duality, is an on-shell symmetry which
relates different Lagrangians associated  to different choices
of symplectic frames.

The only known deformation of ${\cal N}=4$ supergravity compatible with
supersymmetry is the \emph{gauging}. This consists in transmuting
part of the $U(1)^{6+n}$ local symmetry into an non-Abelian gauge
symmetry, or equivalently in promoting part of the global
U-duality symmetry to a local symmetry, using some of the
available vectors. This operation should not alter the total number
of propagating degrees of freedom, as required e.g.~by
supersymmetry.

It is possible to parameterize all gaugings of ${\cal N}=4$ supergravity
by using the so-called \emph{embedding tensor}. The latter describes how the
gauge algebra is realized in terms of the U-duality generators. It
captures all possible situations, including those where some of the gauge
fields are the magnetic duals of the vectors originally present in
the Lagrangian, as well as the option of gauging the duality
rotation between electric and magnetic vectors, which, as already
stressed, appears only as an on-shell symmetry. In this procedure,
one doubles the number of vector degrees of freedom, keeping
however unchanged the number of propagating ones thanks to
appropriate auxiliary fields. One therefore avoids any choice of
symplectic frame until a specific gauging is performed. We will
not elaborate on these general properties but summarize the
structure of the embedding tensor that we will use in the ensuing.
We refer the reader to
\cite{deWit:2005ub,Schon:2006kz,Derendinger:2006jb} for further
information on this subject.

The $n+6$ electric vector fields $\mathbf{A}^{M+},\ M=1,\ldots
,6+n$ belong to the fundamental vector representation of
$SO(6,n)$. Their magnetic duals $\mathbf{A}^{M-}$ form also a
vector of $SO(6,n)$, but carry opposite charge with respect to the
$SO(1,1)\subset SL(2,\mathbb{R})$ that generates the axionic
rescalings.

The $SL(2,\mathbb{R})$ algebra is generated by $S_{\alpha\beta} =
S_{\beta\alpha},\ \alpha,\beta\in\{+,-\}$, which obey the following
commutation relations:
\begin{equation}\label{eq:slcom}
  \left[S_{\alpha\beta},S_{\gamma\delta}\right]=-
  \epsilon_{\alpha\gamma}\, S_{\beta\delta} -
  \epsilon_{\beta\delta}\, S_{\alpha\gamma} -
  \epsilon_{\alpha\delta}\, S_{\beta\gamma} -
  \epsilon_{\beta\gamma}\, S_{\alpha\delta}
\end{equation}
with $\epsilon^{+-}=1=\epsilon_{+-}$. In the vector representation
$\left(S_{\gamma\delta}\right)_\alpha^{\beta} =
  \epsilon_{\gamma\alpha}\delta^\beta_\delta
+ \epsilon_{\delta\alpha}\delta^\beta_\gamma$ and they explicitly
read\footnote{Since with the present conventions for
$\epsilon_{\alpha\beta}$, $\epsilon_{\gamma\alpha}\,
\epsilon^{\gamma\beta} = \delta_\alpha^\beta$, we can raise and
lower $\alpha$-indices unambiguously as follows: $A_\alpha=
A^\beta \, \epsilon_{\beta\alpha}$ and $B^\alpha =
\epsilon^{\alpha\beta}B_\beta$. This leads to $A_+=-A^-$ and
$A_-=A^+$. In particular, $S_{++} = S^{--}$, $S_{+-} = -S^{+-}$
and $S_{--} = S^{++}$. }
\begin{equation}\label{eq:slvec}
   S_{++} = S^{--} =
    \begin{pmatrix}
    0&0\\
    2&0
    \end{pmatrix},\quad
   S_{+-} = -S^{+-}=
    \begin{pmatrix}
    -1&0\\
    0&1
    \end{pmatrix},\quad
    S_{--} = S^{++} =
    \begin{pmatrix}
    0&-2\\
    0&0
    \end{pmatrix}.
\end{equation}

The axion $a$ and the dilaton $\phi$ form a complex scalar
$\tau=a+i \mathrm{e}^{-2\phi}$, which parameterizes the
${SL(2,\mathbb{R})}/{U(1)}$ coset. We define
\begin{equation}\label{eq:dilmat}
   M_{\alpha\beta} = \frac{1}{\mathrm{Im} \tau}
    \begin{pmatrix}
    |\tau|^2& \mathrm{Re} \tau\\
    \mathrm{Re} \tau&1
    \end{pmatrix}
\end{equation}
and we denote by $M^{\alpha \beta}$ its inverse.
The action of
\begin{equation}
g= \begin{pmatrix}
    a&b\\
    c&d
\end{pmatrix} \in SL(2,\mathbb{R})
\end{equation}
is linear on $M$:
$M\to gMg^{T}$ while it acts on $\tau$ as a Möbius transformation:
$\tau \to \nicefrac{a\tau+b}{c\tau+d}$. Therefore, $S^{++}$
generates the axionic shifts,  $S^{+-}$ the axionic rescalings,
whereas the electric-magnetic duality is generated  by  $S^{--}$.
In this basis,
$\left(\left\{\mathbf{A}^{M+}\right\},\left\{\mathbf{A}^{M-}\right\}\right)$
form a doublet of $SL(2,\mathbb{R})$ with diagonal $S^{+-}$ and
correspondingly the rescaling charges are $+1$ and $-1$.

The $SO(6,n)$ is generated by $T_{MN}= - T_{NM}$, $M,N \in \{1,
\ldots, 6+n\}$ obeying
\begin{equation}\label{eq:socom}
\left[T_{KL},T_{JM}\right]=\eta_{LJ} T_{KM}+\eta_{KM}
T_{LJ}-\eta_{KJ} T_{LM} - \eta_{LM} T_{KJ},
\end{equation}
with $\eta_{LJ}$ being the $SO(6,n)$ metric\footnote{Indices $M,N,
\ldots$ are lowered and raised with $\eta_{LJ}$ and $\eta^{KM}$
(inverse matrix).}. In the fundamental representation the
generators read: $\left(T_{KL}\right)_I^{\hphantom{I}J} =
\eta_{KI}\delta^{\hphantom{I}J}_L -
\eta_{LI}\delta^{\hphantom{I}J}_K$.

The $6n$ scalars coming from the $n$ vector multiplets live on the
$SO(6,n)/ (SO(6) \times SO(n))$ coset and can be parameterized by
a symmetric matrix $\mathbf{M}$ of elements $M_{MN}$. Introducing
vielbeins $\mathbf{V}=\left(V_M^m, V_M^a\right)$ with
$m=1,\ldots,6$ and $a=1,\ldots,n$ we can write
$\mathbf{M}=\mathbf{VV}^T$ and define the fully antisymmetric
tensor
\begin{equation}
M_{MNPQRS}=\epsilon_{mnpqrs} V^m_M V^n_N V^p_P V^q_Q V^r_R V^s_S
\end{equation}
that appears in the scalar potential. We will denote by $M^{MN}$
the inverse of $M_{MN}$.

The gauging of ${\cal N}=4$ supergravity proceeds by selecting a
subalgebra of $SL(2,\mathbb{R})\times SO(6,n)$ generated by some
linear combination of $\left\{S_{\alpha\beta}, T_{KL} \right\}$.
The coefficients of this combination are the components of the
embedding tensor and are subject to various constraints discussed
in detail in the aforementioned references.  In summary, this tensor
belongs \textit{a priori} to the $(\mathbf{2}\times \mathbf{3},
\mathbf{Vec} ) + (\mathbf{2}, \mathbf{Vec} \times \mathbf{Adj})$
of $SL(2,\mathbb{R})\times SO(6,n)$. However, there are linear
constraints resulting from the requirement that the commutator
provides an adjoint action and that supersymmetry is
preserved\footnote{This is actually the minimal set of  constraints that
one can consistently impose and they also guarantee that a
Lagrangian exists propagating the correct number of degrees of
freedom, as one learns from general studies on gaugings of maximal
supergravities \cite{deWit:2002vt, deWit:2003hq, deWit:2003hr,deWit:2004nw,deWit:2005ub}.}.
These reduce the representation
content of the embedding
tensor to $(\mathbf{2}, \mathbf{Vec} ) + (\mathbf{2},
\mathbf{Ant}_{[3]})$. Furthermore, there are quadratic constraints
guaranteeing  the closure of the gauge algebra -- Jacobi identity. Putting
everything together, one finds that the admissible generators of the
gauge algebra are of the form
\begin{equation}\label{eq:gaualg}
 \Xi_{\alpha L} = \frac{1}{2} \left( f_{\alpha LMN}\, T^{MN} -\eta^{PQ}\,\xi_{\alpha P} \,
 T_{QL}+
\epsilon^{\gamma\beta}\,\xi_{\beta L}\,
S_{\gamma\alpha}\right)
\end{equation}
where $ f_{\alpha LMN}$ are fully antisymmetric in $L,M,N$ and
$\xi_{\alpha L}$ satisfy
\begin{equation}\label{eq:conone}
\mathrm{(\romannumeral1)}\quad  \eta^{MN}\,\xi_{\alpha M}\,
\xi_{\beta N}=0 \quad \forall \alpha,\beta.
\end{equation}
These parameters characterize completely the gauging.

The  set of
consistency conditions is completed as follows:
\begin{eqnarray}
\mathrm{(\romannumeral2)}&\quad &\eta^{MN}\,\xi_{(\alpha M}\,
f_{\beta) NIJ}=0, \label{eq:contwo}\\
\mathrm{(\romannumeral3)}&\quad &\epsilon^{\alpha\beta}\left(\xi_{
\alpha I }\, \xi_{\beta J} + \eta^{MN}\,\xi_{\alpha M} \,f_{\beta
N I J} \right)=0 \label{eq:conthree}
\end{eqnarray}
and
\begin{eqnarray}
\mathrm{(\romannumeral4)}\quad \eta^{MN}\, f_{\alpha MI[J}\,
f_{\beta KL]N}&=& \frac{1}{2}\xi_{\alpha [J}\,f_{\beta KL]I} +
\frac{1}{6}\epsilon_{\alpha\beta}\, \epsilon^{\gamma\delta}\,
\xi_{\gamma I}\, f_{\delta JKL}\nonumber \\
&-&\frac{1}{2}\eta^{MN}\, \xi_{\alpha M}\, f_{\beta
N[JK}\,\eta_{L]I} - \frac{1}{6}f_{\alpha JKL}\, \xi_{\beta I}
,\label{eq:confour}
\end{eqnarray}
where $[]$ and $()$ stand for antisymmetrization and
symmetrization with respect to different indices belonging to the
same family (e.g. $[L\beta N] = \nicefrac{1}{2}(L\beta N - N\beta
L)$).

Several comments are in order here. A general gauging is
manifestly expressed in terms of $2\times (6+n) + 2 \times
\nicefrac{(6+n)(5+n)(4+n)}{6}$ parameters subject to four
conditions
(\romannumeral1,\romannumeral2,\romannumeral3,\romannumeral4). The
gauge algebra, as defined by this set of parameters, is
characterized by the commutation relations of the subset of
independent $\Xi_{\alpha L}$'s. These generators are indeed
constrained (they satisfy e.g. $\Xi_{(\alpha L}\, \xi_{\beta)M}\,
\eta^{LM}=0$ as a consequence of their definition
(\ref{eq:gaualg}) and Eqs. (\ref{eq:conone}), (\ref{eq:contwo}),
(\ref{eq:conthree})) as they should since no more that $6+n$
vectors can propagate. The structure constants of the gauge
algebra are not directly read off from the  $f_{\alpha LMN}$'s,
which are not necessarily structure constants of some algebra.
They can, however, be expressed in terms of the $f_{\alpha LMN}$'s
and $\xi_{\alpha M}$'s and describe a variety of situations which
capture simple, semi-simple or even non-semi-simple examples. For
all of these, $\eta_{MN}$ always provides an invariant metric
although the Cartan--Killing metric of the corresponding gauge
algebra can be degenerate.

The duality phases of de Roo and Wagemans \cite{deRoo:1985jh,
Wagemans:1987zy, Wagemans:1990mv} are
also captured by the present formalism when the $\xi_{\alpha M}$'s
are absent\footnote{For vanishing $\xi_{\alpha M}$'s, Eq.
(\ref{eq:confour}) is the only constraint; it is an ordinary
Jacobi identity when $\alpha=\beta$.}, as relative orientations of
$f_{+LMN}$ with respect to $f_{- LMN}$ for each simple component
\cite{Schon:2006kz}. We will not elaborate any longer on the
general aspects of the embedding tensor and the variety of
physical possibilities and for further details we will refer the
reader to the already quoted literature.

\subsection{Lagrangian formulation}\label{sec:gengauglag}

The \emph{bosonic} Lagrangian corresponding to the most general
${\cal N}=4$ gauging was presented in \cite{Schon:2006kz}. For the
sake of completeness we reproduce this result here and provide
several comments. The Lagrangian consists of a kinetic term, a
topological term, and a potential for the scalars:
\begin{equation}\label{eq:boslag}
  \mathcal{L} =\mathcal{L}_{\mathrm{kin}} + \mathcal{L}_{\mathrm{top}} + \mathcal{L}_{\mathrm{pot}}.
\end{equation}

The kinetic term reads
\begin{eqnarray}
e^{-1} {\cal L}_{\mathrm{kin}}=\frac{1}{2} R &+& \frac{1}{16}
D_\mu M_{MN}\, D^\mu M^{MN} -\frac{1}{4({\rm Im}\tau)^2} D_\mu
\tau \overline{D^\mu \tau}
 \nonumber\\
&-&\frac{1}{4} \mathrm{e}^{-2\phi} M_{MN}\, H^{M}_{\mu\nu}\,
H^{N\mu\nu} + \frac{1}{8} a \, \eta_{MN}\,
\epsilon^{\mu\nu\kappa\lambda} \,H_{\mu\nu}^M\,
H_{\kappa\lambda}^N
\end{eqnarray}
with the covariant derivatives defined as
\begin{eqnarray}
D_\mu M_{MN}&=&\partial_\mu M_{MN} + 2 g \, A^{P\alpha}_\mu\,
\Theta_{\alpha P(M}^{\hphantom{\alpha P(M}Q}\, M^{\vphantom{Q}}_{N)Q},\\
D_\mu \tau &=&\partial_\mu \tau + g A_\mu^{M-}\, \xi_{+M} + g
\left(A_\mu^{M+}\, \xi_{+M} -A_\mu^{M-}\, \xi_{-M}\right)\tau - i
g\, A_\mu^{M+}\, \xi_{-M}\,  \bar \tau^2
\end{eqnarray}
and the generalized gauge-field strengths being
\begin{equation}
H^M_{\mu\nu}=2 \partial^{\vphantom M}_{[\mu} A_{\nu]}^{M+}- g \hat
f_{\alpha NP}^{\hphantom{\alpha NP} M}\,  A_{[\mu}^{N\alpha}\,
A_{\nu]}^{P+}+\frac{g}{2}
\Theta_{-\hphantom{M}NP}^{\hphantom{-}M}\,
C_{\mu\nu}^{NP}+\frac{g}{2} \xi_{+}^{\hphantom{+}M}\,
C^{++}_{\mu\nu} + \xi_{-}^{\hphantom{+}M} \, C^{+-}_{\mu\nu} .
\end{equation}
The combinations $\Theta_{\alpha MNP}$ and $\hat f_{\alpha MNP}$ are defined
in terms of the gauging parameters $f_{\alpha MNP}$ and $\xi_{\alpha N}$ as
\begin{eqnarray}
\Theta_{\alpha MNP}&=&f_{\alpha MNP}-\xi_{\alpha [ N}\,  \eta_{P]M}, \\
\hat f_{\alpha MNP} &=& f_{\alpha MNP} - \xi_{\alpha [M}\,
\eta_{P]N} -\frac{3}{2} \xi_{\alpha N} \, \eta_{MP}.
\end{eqnarray}

The tensor gauge fields $C_{\mu\nu}^{MN}=C_{\mu\nu}^{[MN]}$ and
$C_{\mu\nu}^{\alpha\beta}=C_{\mu\nu}^{(\alpha\beta)}$ are
auxiliary and their elimination ensures that the correct number of
gauge field degrees of freedom are propagated. For that purpose
one needs to introduce a topological term in the Lagrangian
\begin{eqnarray}
e^{-1} {\cal L}_{\mathrm{top}}&=&-\frac{g}{2}
\epsilon^{\mu\nu\rho\lambda} \Bigg(\xi_{+M} \, \eta_{NP}\,
A^{M-}_\mu  A^{N+}_\nu  \partial_\rho A^{P+}_\lambda -\left(\hat
f_{-MNP}+2 \xi_{-N} \, \eta_{MP}\right) A_\mu^{M-}
A_\nu^{N+} \partial_\rho A_\lambda^{P-}\nonumber\\
&&-\frac{g}{4} \hat f_{\alpha MNR}\,  \hat f_{\beta
PQ}^{\hphantom{\beta PQ}R} A_\mu^{M\alpha} A_\nu^{N+}
A_\rho^{P\beta} A_{\lambda}^{Q-}+ \frac{g}{16}
\Theta_{+MNP}^{\hphantom{M}}\,
\Theta_{-\hphantom{M}QR}^{\hphantom{-}M} \, C_{\mu\nu}^{NP}\,
C_{\rho\lambda}^{QR}\\
&&-\frac{1}{4}\left(\Theta_{-MNP}\,
C_{\mu\nu}^{NP}+\xi_{-M}\,C^{+-}_{\mu\nu} +\xi_{+M}\,
C^{++}_{\mu\nu}\right)\left(2 \partial_\rho A_\lambda^{M-}-g \hat
f_{\alpha QR}^{\hphantom{\alpha QR}M}\, A_\rho^{Q\alpha}\,
A_\lambda^{R-}\right)\Bigg).\nonumber
\end{eqnarray}

Finally, there is a potential for the scalar fields that takes the form
\begin{eqnarray}
e^{-1} {\cal L}_{\mathrm{pot}} &=&-\frac{g^2}{16} \Bigg(f_{\alpha
MNP}\, f_{\beta QRS}\, M^{\alpha\beta}\left(\frac{1}{3} M^{MQ}\,
M^{NR}\, M^{PS} +
\left(\frac{2}{3} \eta^{MQ} - M^{MQ}\right)\eta^{NR} \eta^{PS}\right)\nonumber\\
&&-\frac{4}{9} f_{\alpha MNP} \, f_{\beta QRS}\,
\epsilon^{\alpha\beta} \, M^{MNPQRS} + 3 \xi_{\alpha}^M\,
\xi_{\beta}^N\, M^{\alpha \beta} M_{MN} \Bigg).
\end{eqnarray}

The basic feature of this Lagrangian is that it depends explicitly
on both the electric gauge potentials $A^{M+}_\mu$ and their
magnetic duals $A^{M-}_\mu$. Therefore it allows the gauging of
any subgroup of the full duality group $SL(2,\mathbb{R})\times
SO(6,n)$, where in principle both electric and magnetic potentials
can participate. The field equations derived from this Lagrangian
and the gauge transformations can be found in \cite{Schon:2006kz}.

\section{\boldmath Gauging the ${\cal N}=4$ axionic symmetries\unboldmath }\label{sec:elecgaug}

Here we present a class of ${\cal N}=4$ gauged supergravities
obtained by making local the axionic shifts and axionic
rescalings. This is a subclass of the \emph{electric gaugings} and
we will refer to them as \emph{non-unimodular gaugings}. Their
existence was pointed out in \cite{Schon:2006kz} but here we
analyze in detail and full generality the properties of the
corresponding gauge algebra and discuss some interesting features
of their Lagrangian description, such as a duality
between massive vectors and massive two-forms.

\subsection{Electric-magnetic duality and electric gaugings}\label{sec:elecgaugdef}

The most general ${\cal N}=4$ gauging is described in terms of two tensors
$f_{\alpha LMN}$ and $\xi_{\beta I}$, $\alpha,\beta \in\{+,-\}$
and $I,L,M,N\in\{1,\ldots,6+n\}$, satisfying four quadratic
conditions (\romannumeral1, \romannumeral2, \romannumeral3,
\romannumeral4) displayed in Eqs.~(\ref{eq:conone})--(\ref{eq:confour}). This general formalism,
defined in an arbitrary symplectic frame, captures in particular
the gauging of the electric-magnetic duality symmetry generated by
$S_{++}$ (see Sec. \ref{sec:gengaugemb}).

Gaugings with pure $f_{\alpha LMN}$'s have been studied
extensively in the literature. They correspond to switching on
gauge algebras entirely embedded in $SO(6,n)$, as shown by
(\ref{eq:gaualg}). On the other hand,  turning on  the $\xi_{\beta
I}$'s allows one to gauge both $SL(2,\mathbb{R})$ and $SO(6,n)$.
This situation has not attracted much attention and only a few
examples of the corresponding gauge algebras have been analyzed
(see e.g. \cite{Schon:2006kz,Villadoro:2004ci}). Our aim is to
study systematically a class of such gaugings and show that they
correspond to a specific pattern of higher-dimensional reduction
which generalizes the Scherk--Schwarz mechanism.

We will focus here on \emph{electric gaugings,} namely
gaugings that do not involve the $S_{++}$ generator of
$SL(2,\mathbf{R})$. Axionic shifts $S_{--}$ or axionic rescalings
$S_{-+}$ will however be gauged, accompanied by the appropriate
$SO(6,n)$ generators. Hence, this class of gaugings is defined by setting
\begin{equation}\label{elgau}
\xi_{-I}=0.
\end{equation}
We will also set
\begin{equation}\label{extelgau}
f_{-LMN}=0,
\end{equation}
although this is not compulsory for electric
gaugings\footnote{What is meant by electric gauging is not
universally set in the literature. In \cite{Schon:2006kz}, for
example, electric gaugings are defined as those with $\xi_{\beta
I}=f_{-LMN}=0$, which is somewhat too restrictive.}, while keeping
$\xi_{+ I}\equiv\xi_{I}\neq 0$ and $f_{+LMN}\equiv f_{LMN}\neq 0$.
Furthermore, we will focus on the case $n=6$, which is related to
pure gravity in ten dimensions, and adopt the off-block-diagonal
$6+6$ metric:
\begin{equation}\label{off-eta}
  \mathbf{\eta} =  \begin{pmatrix}
   0 & \mathbb{I}_6\\
   \mathbb{I}_6 & 0
    \end{pmatrix}.
\end{equation}

The quadratic constraint (\romannumeral3), Eq.
(\ref{eq:conthree}), is now automatically satisfied while the constraints
(\romannumeral1, \romannumeral2, \romannumeral4) -- Eqs.~(\ref{eq:conone}), (\ref{eq:contwo}), (\ref{eq:confour}) -- reduce
to
\begin{eqnarray}
&\eta^{MN}\,\xi_{M}\, \xi_{N}=0, \quad \eta^{MN}\,\xi_{M}\,
f_{NIJ}=0,& \label{eq:cononetwoel}\\
&\eta^{MN}\, f_{MI[J}\, f_{KL]N}= \frac{2}{3}f_{[IJK}\,\xi_{L]}.&
\label{eq:confourel}
\end{eqnarray}

\subsection{Non-unimodular gaugings
}\label{sec:elecgaugsol}

The fundamental representation {\bf 12} of $SO(6,6)$ decomposes
into ${\bf 6}_{+1}+{\bf 6}_{-1}$ under the diagonal $GL(6)=U(1) \times SL(6)$ subgroup 
and correspondingly the $I$-indices decompose into $(i,i')$ with both $i$ and $i'$
ranging from $1$ to $6$. Then, the $6+6$ metric can be written in the following
way: $\eta_{ij}= \eta_{i'j'} =0$, whereas $\eta_{ii'}=
\eta_{i'i}= \delta_{ii'}$. In this basis the $SO(6,6)$-invariant
inner product takes the form $A_M B^M = A_m B^m + A_{m'} B^{m'}$
and we can write $A^m\equiv A_{m'}, A^{m'}\equiv A_m$.

\paragraph{A specific solution with $f$'s and $\xi$'s.} Now, a non-trivial solution
to Eqs.~(\ref{eq:cononetwoel}) and (\ref{eq:confourel}) is
\begin{eqnarray}
&\xi_i= \lambda_i \quad \xi_{i'}=0\quad \mathrm{for \ all} \quad i',& \label{eq:solelx}\\
&f_{i'ij}= f_{iji'}=f_{ji'i} =- \lambda_{[i}\, \delta_{j]i'}\quad
\mathrm{all \ others \ vanishing,}& \label{eq:solelf}
\end{eqnarray}
with $\lambda_i$ arbitrary real numbers.
The existence of the gauging described in Eqs.~(\ref{eq:solelx}) and
(\ref{eq:solelf}) was pointed out in Ref. \cite{Schon:2006kz}. Actually,
there exist a whole class of gaugings of this type with more
components of the tensor $f_{IJK}$ turned on. As discussed in  \cite{Schon:2006kz},
besides Eqs.~(\ref{eq:solelx}) and
(\ref{eq:solelf}) one can turn on $f_{ijk}$. Then,  the quadratic constraints
reduce to a single equation
\begin{equation}
f_{[ijk}\lambda_{l]}=0.
\label{bian}
\end{equation}
Since the novel feature of this class of gaugings is the presence
of a non-zero $\xi$ parameter, we will restrict ourselves to the
simplest example, namely the gauging with $f_{ijk}=0$.

The gauging under consideration will be called ``non-unimodular''
for reasons that will become clear at the end of
Sec.~\ref{sec:cogau}, or ``traceful'' since
\begin{equation}\label{eq:trfu}
f_{ij}^{\hphantom{ij}j}=-\frac{5}{2}\lambda_{i}^{\hphantom{j}}.
\end{equation}
This is slightly misleading, however, since the gauge algebra
\emph{is} traceless as a consequence of the full antisymmetry of
its genuine structure constants. The latter \emph{are not}
$f_{ij}^{\hphantom{ij}k}$ but specific combinations of $f_{IJK}$
and $\xi_I$ read off from the commutation relations of generators
(\ref{eq:gaualg}).

\paragraph{The gauge algebra.}In the rest of this section, we will characterize the gauge
algebra, which will be further studied from a higher-dimensional
perspective in the next chapters. Using Eqs.~(\ref{eq:solelx}) and
(\ref{eq:solelf}) in expression (\ref{eq:gaualg}), we obtain:
\begin{eqnarray}
\Xi_{-i}&=& - \frac{\lambda_{i}}{2}S_{--}\equiv  \lambda_{i} \Upsilon, \label{eq:alelxmi}\\
\Xi_{-i'}&=&0,\\
\Xi_{+i}&=&- \frac{\lambda_{i}}{2}\left( T_{\hphantom{j}j}^{j} +
S_{-+} \right)\equiv \lambda_i \Xi,\\
\Xi_{+i'}^{\vphantom{j}}&=&-\lambda_j^{\vphantom{j}}
T_{\hphantom{j}i'}^{j}\equiv \Xi_{i'}^{\vphantom{j}}.
\end{eqnarray}
The gauge algebra at hand has at most 18 non-vanishing generators but
only 7 are independent: $\Upsilon,\Xi$, plus 5 of the $\Xi_{i'}$ 
due to the constraint $\lambda^{i'} \Xi_{i'}=0$.

Their
commutation relations follow from Eqs. (\ref{eq:slcom}) and
(\ref{eq:socom}):
\begin{eqnarray}
\left[\Xi_{i'}, \Xi_{j'}\right]&=& 0,\\
\left[\Upsilon, \Xi_{j'}\right]&=& 0,\\
\left[\Xi_{i'}, \Xi \right]&=& \Xi_{i'},\\
\left[\Upsilon,\Xi\right]&=& - \Upsilon\label{eq:alelxiy}.
\end{eqnarray}
The set $\{\Upsilon,\Xi_{i'}\}$ spans an Abelian
Lie subalgebra. Furthermore, the algebra generated by $\{
\Upsilon, \Xi \}$ is a non-compact $A_{2,2}$ subalgebra of
$SL(2,\mathbb{R})\times SO(6,6)$.

The above algebra is \emph{non-flat}\footnote{An algebra is called flat when
it is generated by a set $\{Q,X_i\}$ satisfying the commutation
relations $[Q,X_i]=M_i^j X_j, [X_i,X_j]=0$ with $M_i^j=-M_j^i$.
Then the Levi--Civita connection on the corresponding group manifold has zero
curvature, therefore justifying the name flat.},
 in contrast to the algebras
obtained by standard Scherk--Schwarz reductions. As we will see in
later, the gaugings at hand are related to twisted versions of
these reductions, which relax therefore the flatness of the gauge
algebras. The particular case $\lambda_i = \lambda
\delta_{1i}$ 
appears when compactifying from five to
four dimensions (see \cite{Villadoro:2004ci}) with a non-compact
twist generated by the five-dimensional rescaling. The full
algebra (\ref{eq:alelxmi})--(\ref{eq:alelxiy}) will also emerge
(see Sec.~\ref{sec:hetered}) as a ten-dimensional heterotic
reduction with twist. Richer non-Abelian extensions are possible
in this case, that eventually lead to non-vanishing $f_{ijk}$ as
already advertised, and which are not possible when compactifying
from five dimensions. These issues will be extensively analyzed in
Sec.~\ref{sec:cogau}.

\subsection{Lagrangian description}
It is straightforward to derive the bosonic Lagrangian for the
gaugings we have just described by using the general formulas of
the previous section. As a first step, we will implement
(\ref{elgau}) and (\ref{extelgau}), and later set
(\ref{eq:solelx}) and (\ref{eq:solelf}). Finally, we will dualize
a vector, which acquires a Stückelberg-like mass via the gauging
at hand, into a massive two-form field potential.

\paragraph{\boldmath Electric gaugings $\xi_{-I}= f_{-KLM}=0$. \unboldmath}

The kinetic terms read
\begin{eqnarray}
e^{-1} {\cal L}_{\mathrm{kin}}=\frac{1}{2} R &+& \frac{1}{16}
D_\mu M_{MN} \, D^\mu M^{MN}
- \frac{1}{4} \mathrm{e}^{4 \phi} D_\mu a \, D^\mu a - D_\mu \, \phi D^\mu \phi \nonumber\\
&-&\frac{1}{4} \mathrm{e}^{-2\phi} M_{MN}\,  H^M_{\mu\nu}\,
H^{N\mu\nu} + \frac{1}{8} a\,  \eta_{MN}\,
\epsilon^{\mu\nu\kappa\lambda} \, H_{\mu\nu}^M\,
H_{\kappa\lambda}^N, \label{n4kin}
\end{eqnarray}
where now
\begin{equation}
D_\mu M_{MN}=\partial_\mu M_{MN} + 2 g \, A^P_\mu\,
\Theta_{P(M}^{\hphantom{P(M}Q} \, M_{N)Q}^{\vphantom{Q}}
\end{equation}
and
\begin{equation}
H_{\mu \nu} ^{M}=2 \partial_{[\mu}^{\vphantom{Q}} A_{\nu]}^M -g
\hat f_{NP}^{\hphantom{NP}M} A_{[\mu}^{N} \, A_{\nu]}^{P}
+\frac{g}{2} \xi^M \, C_{\mu\nu}
\end{equation}
with $C_{\mu\nu}:=C^{++}_{\mu\nu}$.

Since $f_{-MNP}$ and $\xi_{-M}$ are zero and hence $\Theta_{-MNP}$
and $\hat f_{-MNP}$ are zero as well, we have omitted the ``+"
index from all coefficients.  We will use the notation
$A_\mu^M\equiv A_\mu^{M+}, A_\mu^{M-}\equiv X^M_{\mu}$ for the
gauge potentials in order to avoid cluttering of the formulas with
indices. Furthermore, we define the linear combinations
\begin{equation}
  X_\mu\equiv\xi_M X^M_\mu , \quad A_\mu\equiv\xi_M
A^M_\mu.
\end{equation}
The gauge covariant derivatives of the axion and dilaton take the
form\footnote{From now on we set the coupling constant $g$ equal
to 1 for simplicity.}
\begin{eqnarray}
D_\mu a &=&\partial_\mu a+  X_\mu + A_\mu a, \\
D_\mu \phi &=& \partial_\mu \phi -\frac{1}{2}  A_\mu.
\end{eqnarray}
By turning on the parameters $\xi$  we have gauged a non-Abelian
two-dimensional subgroup of the $SL(2,\mathbb{R})$ global
axion-dilaton symmetry. The ``magnetic" potential $X_\mu$
corresponds to the gauging of the shift symmetry of the axion $a
\rightarrow a +c$ and acts as a Stückelberg field, while $A_\mu$
gauges the dilatation symmetry $a \rightarrow \mathrm{e}^{-2
\lambda } a$, $\phi\rightarrow \phi + \lambda$. In terms of the
notation of the previous subsection, the corresponding gauge
algebra is the one spanned by $\Xi$ and $\Upsilon$ with
commutation relation (\ref{eq:alelxiy}).

The topological term in the Lagrangian for the class of gaugings
under consideration becomes
\begin{eqnarray}\label{n4top}
e^{-1} \mathcal{L}_{\mathrm{top}}=-\frac{1}{2}
\epsilon^{\mu\nu\rho\lambda} \Big(\xi_M \, \eta_{NP}\,  X^M_\mu
A^N_\nu \partial_\rho A_\lambda^P  &-&\frac{1}{4}
\xi_M^{\vphantom{M}} C_{\mu\nu}^{\vphantom{M}} G^M_{\mu\nu}
\nonumber\\ &-& \frac{1}{4} \hat f_{MNR}^{\vphantom{M}} \,\hat
f_{PQ}^{\hphantom{PQ}R} \, A^M_\mu A^N_\nu A^P_\rho
X^Q_\lambda\Big).
\end{eqnarray}
We have defined the following gauge field strengths
\begin{eqnarray}
F_{\mu\nu}^M&=&2 \partial_{[\mu}^{\vphantom{M}} A_{\nu]}^M
- \hat f_{NP}^{\hphantom{NP}M} A_{[\mu}^{N} A_{\nu]}^{P},\\
G_{\mu\nu}^M&=&2 \partial_{[\kappa}^{\vphantom{M}}
X_{\lambda]}^M-\hat f_{QR}^{\hphantom{QR}M} A^{Q}_{[\kappa}
X^{R}_{\lambda]},
\end{eqnarray}
in terms of which one can write
$H_{\mu\nu}^M=F_{\mu\nu}^M+\frac{1}{2} \xi^M C_{\mu\nu}$ .

Finally, the potential terms are
\begin{eqnarray}
e^{-1}{\cal L}_{\mathrm{pot}}&=& -\frac{1}{16}  M^{++} \bigg(3
\xi_M \xi_N M^{MN}
\nonumber\\
&&+ f_{MNP} f_{QRS} \left(\frac{1}{3} M^{MQ} M^{NR} M^{PS}
+\left(\frac{2}{3}\eta^{MQ}-M^{MQ}\right) \eta^{NR}
\eta^{PS}\right)\bigg) \label{n4pot}
\end{eqnarray}
with $M^{++}=\mathrm{e}^{2\phi}$.

\paragraph{Non-unimodular gaugings.}
We will now specify the gauge parameters by setting
(\ref{eq:solelx}) and (\ref{eq:solelf}). With this gauging, the
indices $i$ and $i'$ corresponding to the ${\bf 6}_{+1}$ and ${\bf 6}_{-1}$
of $U(1) \times SL(6) \subset SO(6,6)$ are treated differently. We first
notice that the magnetic  field strengths $G^{m'}_{\mu\nu}$ do not
appear in the Lagrangian. The gauge field strengths that appear
are
\begin{eqnarray}
F^m_{\mu\nu}&=&2 \partial_{[\mu}^{\vphantom{m}} A_{\nu]}^m\label{gfsone},\\
F^{m'}_{\mu\nu}&=&2\partial_{[\mu}^{\vphantom{m}}
A_{\nu]}^{m'}+\left(2\lambda_j A^{m'}_{[\mu}
A_{\nu]}^j-\lambda^{m'} A_{j[\mu}^{\vphantom{m}}A^j_{\nu]}\right),\label{gfstwo}\\
G^m_{\kappa\lambda}&=&2 \partial_{[\kappa}^{\vphantom{m}}
X_{\lambda]}^m +  \left(\lambda_i A^i_{[\kappa} X^m_{\lambda]}+
\lambda_i A^m_{[\kappa} X_{\lambda]}^i\right),
\end{eqnarray}
where now
\begin{equation}
  X_\mu\equiv\xi_M X^M_\mu=\lambda_m X^m_\mu, \quad A_\mu\equiv\xi_M
A^M_\mu=\lambda_m A^m_\mu.
\end{equation}
One further notices that the Lagrangian actually depends only on
the linear combination $\lambda_m G^m_{\mu\nu}$.
This combination can
be written in terms of $A_\mu$ and $X_\mu$ as
\begin{equation}
 G_{\mu\nu}=2\partial_{[\mu} X_{\nu]} + 2 A_{[\mu} X_{\nu]}.
 \end{equation}
Similarly we introduce $F_{\mu\nu}=\lambda_m F^m_{\mu\nu}=2
\partial_{[\mu} A_{\nu]}$.

The natural prescription is to integrate out the auxiliary
two-form $C_{\mu\nu}$ in order to obtain the final Lagrangian. If
we do so, starting from (\ref{n4kin}), (\ref{n4top}) and
(\ref{n4pot}), we obtain
\begin{eqnarray}
e^{-1}{\cal L}=\frac{1}{2} R &+& \frac{1}{16} D_\mu M_{MN}\, D^\mu
M^{MN}
- \frac{1}{4} \mathrm{e}^{4 \phi}\, D_\mu a\, D^\mu a - D_\mu \phi \,D^\mu \phi \nonumber\\
&-&\frac{1}{4} \mathrm{e}^{-2\phi} \,M_{MN} \,F^M_{\mu\nu}
F^{N\mu\nu}_{\vphantom{\mu}} + \frac{1}{4} a \, \eta_{MN}\,
F_{\mu\nu}^M\, \tilde F^{N\mu\nu}_{\vphantom{\mu}}
-\frac{1}{4\lambda^2} \mathrm{e}^{2\phi}\,  Z^{\mu\nu}\,
Z_{\mu\nu}
\nonumber\\
&-&\frac{1}{12}  \epsilon^{\mu\nu\rho\lambda}\,  X_\mu\,
\omega_{\rho\lambda\nu}  +e^{-1}  {\cal L}_{\mathrm{pot}}.
\label{finalcompact}
 \end{eqnarray}
In this expression $Z^{\mu\nu} = a\, F^{\mu\nu}+
G^{\mu\nu}+\mathrm{e}^{-2\phi}\,  \lambda^{m'} \, M_{m'N}\, \tilde
F^{N\mu\nu} $, where $\tilde  F^{N\mu\nu}$ is the Hodge--Poincaré
dual of $F^{N}_{\mu\nu}$;  we have also  defined $\lambda^2 =
\lambda_m M^{mn} \lambda_n$ and we have introduced the
Chern--Simons form:
\begin{equation}
\omega_{\rho\lambda\nu}^{\vphantom{n}}=A^n_\nu \, F_{\rho\lambda
n}^{\vphantom{n}} + A_{n\nu}^{\vphantom{n}} \, F^n_{\rho\lambda}
-\frac{1}{2} A_\lambda^{\vphantom{n}}\, A^m_\nu\,
A_{m\rho}^{\vphantom{n}}+ \frac{1}{2} A_\nu^{\vphantom{n}}\,
A^m_\lambda \,A_{m\rho}^{\vphantom{n}} + \mathrm{cyclic}.
\end{equation}

The local gauge invariance under axionic shifts enables us to
gauge away the axion $a=0$. Then, the magnetic potential $X_\mu$
acquires a mass through its Stückelberg coupling to the axion. The
final expression for the gauge-fixed Lagrangian is therefore
\begin{eqnarray}
e^{-1}{\cal L}=\frac{1}{2} R &+& \frac{1}{16} D_\mu M_{MN} \,
D^\mu M^{MN}
 - D_\mu \phi \, D^\mu \phi
-\frac{1}{4} \mathrm{e}^{-2\phi}\,  M_{MN}\,  F^M_{\mu\nu}\,
F^{N\mu\nu}_{\vphantom{\mu}}
\nonumber\\
&-& \frac{1}{4\lambda^2} \mathrm{e}^{2\phi} \left(
G^{\mu\nu}+\mathrm{e}^{-2\phi}\, \lambda^{m'}\,  M_{m'N} \, \tilde
F^{N\mu\nu}\right) \left( G_{\mu\nu}+\mathrm{e}^{-2\phi}\,
\lambda^{m'}\,  M_{m'N} \, \tilde F^{N}_{\mu\nu}\right) \nonumber
\\ &-&\frac{1}{12} \epsilon^{\mu\nu\rho\lambda} \, X_\mu \,
\omega_{\rho\lambda\nu}+e^{-1} {\cal L}_{\mathrm{pot}} .
\label{finalcompactfixed}
 \end{eqnarray}
It captures all the relevant information carried by the
axionic-symmetry gauging.

\paragraph{Stückelberg mass and dualization.}
There is a different formulation of the theory, which is actually
more suggestive of a higher-dimensional origin. Instead of
integrating out the auxiliary antisymmetric tensor $C_{\mu\nu}$,
one can promote $C_{\mu\nu}$ to a propagating field and integrate
out $X_\mu$. Using as previously (\ref{n4kin}), (\ref{n4top}) and
(\ref{n4pot}), this procedure yields the following Lagrangian:
\begin{eqnarray}
e^{-1}\tilde{ {\cal L}}=\frac{1}{2} R &+& \frac{1}{16} D_\mu
M_{MN}
\,D^\mu M^{MN} - D_\mu \phi \, D^\mu \phi \nonumber\\
&-&\frac{1}{4} \mathrm{e}^{-2\phi} \, M_{MN}
\left(F_{\mu\nu}^M+\frac{1}{2} \xi^M\, C_{\mu\nu}\right)
\left(F^{N \mu\nu}+\frac{1}{2} \xi^N\, C^{\mu\nu}\right)\nonumber\\
&-&\frac{3}{8} \mathrm{e}^{-4 \phi} \left(\partial_{[\kappa}\,
C_{\mu\nu]}-A_{[\kappa} \,C_{\mu\nu]}- \frac{1}{3}
\omega_{\kappa\mu\nu} \right)^2+e^{-1} {\cal L}_{\mathrm{pot}} ,
\label{gsaux}
\end{eqnarray}
where the two-form $C_{\mu\nu}$ acquires now a
scalar-field-dependent mass.

The key observation is that the theory described by (\ref{gsaux})
is related to that described by (\ref{finalcompactfixed}) by an
interesting  generalized duality. Recall that a massless two-form
potential in four dimensions carries one propagating degree of
freedom and can be dualized into a scalar. In the case where the
two-form comes from the reduction of the ten-dimensional NS-NS
two-form in the gravity multiplet, the dual scalar is the axion.
If instead, as it happens here, the two-form is massive, it
carries three degrees of freedom and the dual potential is a
massive vector. This duality is a particular instance of a
generalized duality between massive $p$-forms and massive
($D-p-1$)-forms in $D$ dimensions \cite{Townsend:1983xs, Quevedo:1995ep}.

For the benefit of the reader we will present schematically how
this generalized duality works, leaving as an exercise its precise
implementation between (\ref{finalcompactfixed}) and
(\ref{gsaux}). Start from a massive gauge field $X_\mu$ with
Lagrangian
\begin{equation}
{\cal L}=-\frac{1}{4g^2} \left(\partial_\mu X_\nu -\partial_\nu
X_\mu\right)\left(\partial^\mu X^\nu -\partial^\nu X^\mu\right) +
\frac{1}{2} m^2 X_\mu X^\mu.
\end{equation}
This Lagrangian can be obtained from
\begin{equation}
\tilde{\tilde{{\cal L}}}=2 C_{\mu\nu} \, \partial^\mu X^\nu + g^2
C_{\mu\nu}\, C^{\mu\nu} + \frac{1}{2} m^2 X_\mu X^\mu
\end{equation}
by integrating $C_{\mu\nu}$, which is an auxiliary non-propagating
antisymmetric tensor. Instead, we can integrate by parts
$\tilde{\tilde{{\cal L}}}$ so that $X_\mu$ becomes non-dynamical,
while $C_{\mu\nu}$ acquires a dynamics. Integrating out finally
$X_\mu$ yields an action for a massive two-form,
\begin{equation}
\tilde{{\cal L}}=-\frac{2}{m^2} \partial_\mu C^{\mu\nu}\,
\partial^\lambda C_{\lambda\nu}+ g^2 C_{\mu\nu}\, C^{\mu\nu},
\end{equation}
which can be brought to a more familiar form for antisymmetric
tensors, by Hodge--Poincaré-dualizing  $C_{\mu\nu}$.

Following a similar pattern, one can replace the massive vector
$X_\mu$ in $\mathcal{L}$ (Eq. (\ref{finalcompactfixed})) by a
two-form. This yields precisely $\tilde{\mathcal{L}}$ (Eq.
(\ref{gsaux})), therefore demonstrating that the Lagrangians
$\mathcal{L}$ and $\tilde{\mathcal{L}}$ describe equivalent
physics.

\section{Higher-dimensional origin}\label{sec:hetered}

In this section we perform a generalized dimensional reduction of
heterotic supergravity to four dimensions and show that the
resulting effective theory belongs to the class of ${\cal N}=4$
gauged supergravities studied in the previous section. Recall that
the usual dimensional reduction of heterotic supergravity, which
can be thought of as compactification on a six-torus keeping only
the massless modes, results in a four-dimensional theory with 16
supercharges, Abelian gauge group $U(1)^{12+p}$ and a global
off-shell symmetry $SO(6,6+p)$ \cite{Maharana:1992my}. Six of the
Abelian vectors are graviphotons while another six of them come
from reducing the NS-NS two-form on the 1-cycles of the torus.

One can also reduce some of the vectors present already in ten
dimensions and obtain $p$ additional Abelian gauge fields. Usually
this is done for the vectors lying in the Cartan torus of the
ten-dimensional gauge group, therefore yielding a theory with
$p=16$. Since our goal here is to make contact with a gauged
supergravity with $p=0$ we will ignore this possibility and
consider the reduction of the gravity-multiplet fields only. It is
straightforward to extend the reduction to the Yang--Mills sector
and obtain generalizations of the gaugings we discussed so far.

\subsection{\boldmath Heterotic reduction with duality twist \unboldmath}
\label{sec:hetred}

Our starting point is the bosonic action of heterotic supergravity
in the string frame\footnote{The ten-dimensional spacetime indices
$M,N,\ldots$ of this section should not be confused with the
fundamental $SO(6,6)$ indices of the previous sections.}
\begin{eqnarray}
S= \int_{M_4} \mathrm{d}x \int_{K_6} \mathrm{d}y \,  \sqrt{-G}\,
\mathrm{e}^{-\Phi} \bigg(R&+&G^{MN}\,
\partial_M \Phi\,  \partial_N \Phi\nonumber\\
&-&\frac{1}{12} G^{MM'}G^{NN'}G^{KK'} H_{MNK} H_{M'N'K'} \bigg).
\label{actionstring}
\end{eqnarray}
We assume a decomposition of the ten-dimensional spacetime in a
four-dimensional non-compact part $M_4$  parameterized by
coordinates $x^\mu$ and a six-dimensional internal manifold  $K_6$
parameterized by $y^i$. The spacetime indices will be decomposed
as $M=(\mu,i)$. As usual $\Phi$ is the dilaton and $H=\mathrm{d}B$
is the three-form field strength of the NS-NS antisymmetric tensor
$B$. As said, we neglect the ten-dimensional Yang--Mills fields.

Taking $K_6$ to be a flat six-torus and keeping only the
$y$-independent modes yields a theory with $12$ Abelian vectors
and $38$ massless scalars, two of which are the dilaton and the
axion. The latter is the dual of the two-form $B_{\mu\nu}$
obtained by reducing $B_{MN}$. One way to obtain a more
interesting theory is by introducing fluxes in the torus. The most
general reduction of this type, where both NS-NS and geometric
fluxes were present, was studied in \cite{Kaloper:1999yr}.

The introduction of geometric fluxes has an alternative
interpretation as a reduction with a twist for the spacetime
fields \cite{Scherk:1979zr}. This is actually a particular case of
a generalized reduction scheme  that usually is referred to as
``Scherk--Schwarz'' reduction \cite{Scherk:1978ta}. The
characteristic property of reductions of this type is that the
reduction ansatz  can incorporate a dependance on the coordinates
of the internal torus. This dependance is not arbitrary however;
on a technical level it is dictated by the requirement  that the
Lagrangian should be independent of the internal coordinates. This
is implemented by selecting a profile for the fields whose
consistency is guaranteed by some symmetry of the original theory.
Such reductions in the context of supergravity have been studied
in \cite{Bergshoeff:1996ui,Cowdall:1996tw, Bergshoeff:1997mg,
Lavrinenko:1997qa, Kaloper:1998kr, Hull:1998vy, Hull:2003kr,
Villadoro:2004ci, Reid-Edwards:2006vu} while the reader is
referred to \cite{Dabholkar:2002sy} for a general discussion on
reductions with duality twists.

A subtle point that is not usually emphasized is the following.
These reduction schemes yield an effective theory for a finite set
of modes selected out of the infinitude of higher-dimensional
modes according to some symmetry principle. Hence, it is not
necessarily  true that they encompass all low-energy modes and
further analysis is required in order to establish that the
effective theory obtained through a Scherk--Schwarz reduction is
actually a {\em low-energy} effective theory. It would be interesting
to perform such an analysis for the reduction scheme presented below
but this issue lies beyond the scope of this paper.

The symmetry we will employ in the present paper is the $SO(1,1)$
scaling symmetry of (\ref{actionstring}), under which the fields
transform as
\begin{equation}
\Phi \rightarrow \Phi + 4 \lambda,\quad G_{MN} \rightarrow
\mathrm{e}^{\lambda} G_{MN}(x),\quad B_{MN} \rightarrow
\mathrm{e}^{\lambda} B_{MN}(x).
\end{equation}
This enables us to trade the usual periodic ansatz, which assumes
no dependance on the torus coordinates, with the following one
\begin{equation}
\Phi(x,y)=\Phi(x)+ 4 \lambda_i y^i,\quad
G_{MN}(x,y)=\mathrm{e}^{\lambda_i y^i} G_{MN}(x),\quad
B_{MN}(x,y)=\mathrm{e}^{\lambda_i y^i} B_{MN}(x). \label{ansatzhd}
\end{equation}
The parameters $\lambda_i$ are arbitrary real numbers that dictate
the twisting of the fields along the six one-cycles of the torus.

The decomposition of the  ten-dimensional metric tensor in terms
of four-dimensional fields is the usual one
\begin{equation}
G_{\mu \nu}=g_{\mu\nu} + A_{\mu}^i \, A_{\nu}^j\,  h_{ij},\quad
G_{\mu i}=A_{\mu }^j \, h_{ij},\quad G_{i j}=h_{ij}.
\end{equation}
Here $g_{\mu \nu}$ is the metric on $M_4$, $h_{ij}$ are the 21
metric moduli of the six-torus and $A_\mu^i$ are the Kaluza--Klein
gauge fields. Similarly, the antisymmetric tensor is decomposed as
\begin{equation}
B_{\mu \nu}=b_{\mu \nu},\quad
B_{\mu i}=b_{\mu i},\quad
B_{ij}=b_{ij}.
\end{equation}
We obtain a two-form $b_{\mu\nu}$, which usually is dualized to an
axion, six vectors $B^i_\mu$ and 15 scalar moduli $b_{ij}$.

The above decompositions hold for the full ten-dimensional fields
that are assumed to have a dependance on $y^i$ of the type
dictated by the $SO(1,1)$ scaling symmetry. Consistency implies
that
\begin{eqnarray}
&g_{\mu \nu}(x,y)=\mathrm{e}^{\lambda_k y^k} g_{\mu \nu}(x),\quad
b_{\mu \nu}(x,y)=\mathrm{e}^{\lambda_k y^k} b_{\mu \nu}(x),\quad
h_{i j}(x,y)=\mathrm{e}^{\lambda_k y^k} h_{i j}(x),&\label{eq:gy}\\
&b_{i j}(x,y)=\mathrm{e}^{\lambda_k y^k} b_{i j}(x),\quad A_{\mu
}^i(x,y)=A_{\mu }^i(x),\quad
b_{\mu i}(x,y)=\mathrm{e}^{\lambda_k y^k} b_{\mu i}(x),&\label{eq:gb}\\
&\phi(x,y)=\phi(x)+\lambda_k y^k,\label{eq:phiy}&
\end{eqnarray}
where the four-dimensional dilaton is defined as
$\phi=\Phi-\frac{1}{2}\log \det \mathbf{h}$. The $y$-independent
modes on the right-hand sides are the four-dimensional fields for
which we would like to derive the effective action. The ansatz is
consistent in the sense that the $y$-dependance is totally
eliminated from the action and the integration over $y^i$ yields
an overall multiplicative factor. Notice that the volume of the
internal six-torus is encoded in the metric moduli $h_{ij}$.

Let us first reduce the Einstein--Hilbert part of the action along
with the dilaton kinetic term. The most efficient way of
performing this reduction is the following. We start from the
metric
\begin{equation}
G'_{MN}(x,y)=\Omega^2(y)\,  G_{MN}(x),
\end{equation}
where $\Omega(y)=\exp(\frac{1}{2} y^i \lambda_i)$, use the
relation between Ricci scalars for conformally related metrics,
and finally apply the usual reduction formulas for $G_{MN}(x)$.
After the above redefinition of the dilaton,
necessary for absorbing the factor of $\sqrt{h}$, the conformal
rescaling $g_{\mu\nu} \to \frac{\exp{\phi}}{2} g_{\mu\nu}$, which
brings us to the Einstein frame in four dimensions, and a final
rescaling $\phi \to 2 \phi$, we obtain
\begin{eqnarray}
S_{\mathrm{gravity}}=\int_{M_4} \mathrm{d}x \sqrt{- g}  \bigg(
\frac{1}{2} R_4 &+& \frac{1}{8} \left(\partial_\mu\, h_{ij} -
A_\mu^k \, \lambda_k^{\vphantom{m}} \,
h_{ij}^{\vphantom{m}}\right) \left(\partial^\mu h^{ij} + A^{\ell
\mu} \,\lambda_\ell\, h^{ij}\right)
\nonumber\\
 &-& g^{\mu \nu} \left(\partial_\mu \phi
 -\frac{1}{2} A_\mu^i\, \lambda_i^{\vphantom{m}}\right)
\left(\partial_\nu \phi -\frac{1}{2} A^j_\nu\,
\lambda_j^{\vphantom{m}}\right) \nonumber\\&-&\frac{1}{4}
\mathrm{e}^{-2 \phi}\, h_{ij}\, F^i_{\mu\nu} \,
F^{j\mu\nu}_{\vphantom{m}} - \frac{1}{2} \mathrm{e}^{2 \phi}
\,\lambda_i\,   h^{ij} \,\lambda_j \bigg),\label{grav}
\end{eqnarray}
where $F_{\mu\nu}^m=\partial_\mu^{\vphantom{m}} A_{\nu}^m -
\partial_\nu^{\vphantom{m}} A_{\mu}^m$ and all fields are
exclusively $y$-dependent.

There are several observations in order. First, the metric moduli
$h_{ij}$ become charged under the Kaluza--Klein gauge fields
$A^i_\mu$. The charges are given by the vector of twisting
coefficients $\lambda_i$. The dilaton is also coupled in a
Stückelberg fashion to $A_\mu^i$. This signals the gauging of a
shift symmetry as expected from a reduction where the ansatz was
twisted by employing such a symmetry\footnote{ Note that the
dilaton is shifted under the $SO(1,1)$ scaling.}. Notice, furthermore, that
the twisting does not result in a non-Abelian gauge symmetry for
the Kaluza--Klein gauge fields. It does however lead to a
potential for the dilaton and the metric moduli.

The reduction of the NS-NS part of the Lagrangian is more easily
performed using the tangent-space components of  the antisymmetric
three-form \cite{Maharana:1992my}. Furthermore, some field
redefinitions are necessary in order to bring the resulting
Lagrangian to a more standard form. One defines vector fields
$Y_{\mu n}$ and a two-form ${\cal B}_{\mu\nu}$ as
\begin{eqnarray}
Y_{\mu n}^{\vphantom{[}} & = & b_{\mu n}^{\vphantom{[}} + b_{nm}^{\vphantom{[}}\, A^m_{\mu}, \\
{\cal B}_{\mu \nu}^{\vphantom{[}} & = & b_{\mu \nu}^{\vphantom{[}}
+  A^m_{[\mu} \, Y_{\nu] m}^{\vphantom{[}} -A^m_{\mu} A^n_{\nu}
b_{mn}^{\vphantom{[}} .
\end{eqnarray}
We get
\begin{eqnarray}
S_{\mathrm{NS-NS}}=\int_{M_4}\bigg( &-&\frac{1}{8} h^{mn}\,
h^{\ell k}\, D_{\mu} b_{m\ell} \, D^\mu b_{nk}
\nonumber\\&-&\frac{1}{6} \mathrm{e}^{-4\phi}
\left(3\left(\partial_{[\mu}^{\vphantom{m}} \,{\cal
B}_{\nu\lambda]}^{\vphantom{^m}}-\lambda_k^{\vphantom{m}}\,
A^k_{[\mu} \,{\cal B}_{\nu\lambda]}^{\vphantom{m}}\right)
-\frac{1}{2} \Omega_{\mu \nu
\lambda}^{\vphantom{m}}\right)\nonumber\\
&& \times \left(3\left(\partial^{[\mu}_{\vphantom{m}} \,{\cal
B}^{\nu\lambda]}_{\vphantom{^m}}-\lambda^\ell_{\vphantom{m}}\,
A_\ell^{[\mu} \,{\cal B}^{\nu\lambda]}_{\vphantom{m}}\right)
-\frac{1}{2} \Omega^{\mu \nu \lambda}_{\vphantom{m}}\right)
\label{nsns}
\\
&-&\frac{1}{4}\, \mathrm{e}^{-2\phi} \,h^{mn} \left(Y_{\mu\nu m} +
\lambda_m \,{\cal B}_{\mu\nu}- b_{m\ell}^{\vphantom{m}}\,
F^\ell_{\mu\nu}\right) \left(Y^{\mu\nu}_n + \lambda_n \,{\cal
B}^{\mu\nu}- b_{nk}\, F^{k \mu\nu}\right)
 \nonumber\\
&+& \frac{1}{16}  \mathrm{e}^{2 \phi} \lambda_i \left(  h^{ij}\,
h^{k\ell}\, b_{\ell m}\, h^{mn}\, b_{nk} - 2 h^{ik}\, b_{km}\,
h^{mn}\, b_{nr} \,h^{rj}\right) \lambda_j\bigg) \sqrt{- g}\,
\mathrm{d}x , \nonumber
\end{eqnarray}
where we have defined
\begin{eqnarray}
D_\mu b_{n\ell}&=&\partial_\mu \, b_{n\ell}+\lambda_\ell\,  Y_{\mu
n} - \lambda_n \, Y_{\mu \ell} -
\lambda_m \, A^m_{\mu}\,  b_{n\ell}^{\vphantom{m}} ,\\
Y_{\mu \nu \ell}&=& \partial_\mu Y_{\nu \ell} - \partial_\nu
Y_{\mu \ell} +\left(\frac{1}{2} \lambda_\ell\, \delta^{mn}
-\lambda^m \,\delta^n_\ell \right) \left(A_{\mu m}\, Y_{\nu n} -
Y_{\mu n}\, A_{\nu m}\right),
\end{eqnarray}
while the Chern--Simons three-form is
\begin{equation}
\Omega_{\mu\nu\lambda}= Y_{\mu \nu \ell}\, A^\ell_{\lambda} +
A_{\mu \nu}^\ell \, Y_{\ell\lambda}
 - \frac{1}{2} \lambda_\ell \, A_{\nu}^\ell \, A_{\lambda}^m \, Y_{\mu m}+\frac{1}{2}
\lambda_\ell \, A^\ell_{\lambda}\, A_{\nu}^m \,Y_{\mu m} +
\mathrm{cyclic}.
\end{equation}

We observe that the NS-NS moduli are charged under the
Kaluza--Klein gauge fields but have also Stückelberg couplings to
the NS-NS gauge potentials. The latter couplings are due to the
gauging of the shift symmetries of those moduli induced by the
duality twist. A crucial difference with the case of the ordinary
dimensional reduction is that the four-dimensional two-form ${\cal
B}_{\mu\nu}$ acquires a mass. This prohibits the standard dual
formulation in terms of an axion but, according to the discussion
of the previous section on dualities between massive fields,
suggests that a dual formulation in terms of a massive vector is
possible. Let us finally stress that the reduction of the NS-NS
sector also contributes to the potential for the $h_{ij}$ and
$b_{ij}$ moduli (last line of (\ref{nsns})).

\subsection{Contact with ${\cal N}=4$ gauged supergravity}\label{sec:cogau}

We will now show that the effective theory described by the sum of
actions (\ref{grav}) and (\ref{nsns}) is nothing but the ${\cal
N}=4$ gauged supergravity worked out in Sec.~\ref{sec:elecgaug}.
Using the standard parameterization of the moduli matrix $M^{MN}$
\begin{equation}
M^{MN}=\begin{pmatrix}
    h^{mn} & \ - h^{mk}\, b_{kn} \\
    b_{mk}\, h^{kn}  & \ h_{mn}-b_{mk}\, h^{k\ell} \, b_{\ell n}
  \end{pmatrix},
\end{equation}
the ${\cal N}=4$ potential (\ref{n4pot}) obtained for the
non-unimodular gauging reads:
\begin{equation}
V=\frac{1}{16}  \mathrm{e}^{2 \phi} \lambda_i \left( 8 h^{ij}-
h^{ij}\, h^{k\ell}\, b_{\ell m}\, h^{mn}\, b_{nk} + 2 h^{ik}\,
b_{km}\, h^{mn}\, b_{nr}\, h^{rj}\right) \lambda_j .
\end{equation}
This is precisely the potential in the effective theory
(\ref{grav}) plus (\ref{nsns}). Notice that this identification
clarifies the higher-dimensional interpretation of the gauging
parameters $\xi$: they correspond to the parameters used to twist
the boundary conditions by $SO(1,1)$ scalings along the six
one-cycles of the torus.

It is straightforward to check that the rest of the terms in
(\ref{grav}) plus (\ref{nsns}) match exactly  those of
(\ref{gsaux}) provided we identify the gauge fields $A^m_\mu,
A^{m'}_\mu$ in the gauged-supergravity Lagrangian with the
Kaluza--Klein  and NS-NS gauge fields  $A^m_\mu, Y_{m\mu}$ in the
heterotic reduction and the antisymmetric tensors as $C_{\mu\nu}
\leftrightarrow 2 {\cal B}_{\mu\nu}$. This elucidates the
higher-dimensional origin of the gauged supergravity of
Sec.~\ref{sec:elecgaug} and confirms the prominent role of the
generalized duality performed in four dimensions for reaching
(\ref{gsaux}). It is amusing that the four-dimensional two-form
${\cal B}_{\mu\nu}$ that comes from the NS-NS antisymmetric tensor
in ten dimensions is actually the auxiliary tensor gauge field
required for consistency of the gauging in the formalism of
\cite{deWit:2005ub}.

Let us mention at this point that ordinary reductions of the
heterotic theory with NS-NS fluxes and geometric fluxes also yield
${\cal N}=4$ gauged supergravities \cite{Kaloper:1999yr}. The
correspondence with the embedding-tensor language is as follows:
there are no $\xi$'s turned on and the only non-vanishing
parameters are the $f_{+IJK}\equiv f_{IJK}$. Under the
decomposition of indices $I=(i,i')$, the background NS-NS fluxes
$\beta_{ijk}$ and geometric fluxes\footnote{We use the notation of
\cite{Kaloper:1999yr}.} $\gamma_{ij}^{\hphantom{ij}k}$ are
identified with the components of $f_{IJK}$ as
\begin{equation}
f_{ijk}=-3 \beta_{ijk},\quad f_{ijk'}=2
\gamma_{ij}^{\hphantom{ij}k},
\end{equation}
all other components being zero. The remaining non-trivial
quadratic constraint is (\romannumeral4) (Eq. (\ref{eq:confour}))
and it corresponds to the Bianchi identities for the NS-NS fluxes
and the Jacobi identity for the geometric fluxes. From this we
conclude that the more general class of gaugings we mentioned in
Sec.~\ref{sec:elecgaugsol} with non-zero $f_{ijk}$ originates from
a ten-dimensional reduction with an $SO(1,1)$ duality twist
combined with background NS-NS fluxes. The condition (\ref{bian})
found then is a consequence of the Bianchi identity resulting from
the ansatz (\ref{ansatzhd}).

An interesting observation is in order here. The correspondence
between the components $f_{ijk'}$ and the geometric fluxes
$\gamma_{ij}^{\hphantom{ij}k}$ provides an alternative perspective
on the gauging we have performed in Sec.~\ref{sec:elecgaugsol} and
the subsequent heterotic reduction of Sec.~\ref{sec:hetred}.
Indeed, if we interpret the $y$-dependance of the internal metric
(c.f.~Eqs. (\ref{eq:gy})) as inducing a geometric
flux, this flux is automatically ``non-unimodular" since
$\gamma^{\hphantom{ij}i}_{ij}\neq 0$. In ordinary reductions, the
unimodularity condition ensures consistency of the  truncation of
the higher-dimensional Lagrangian \cite{Scherk:1979zr,
Hull:2005hk}.
This well-known obstruction is circumvented in our
approach thanks to the compensating duality twist\footnote{This 
statement refers to a reduction performed 
in the string frame. The metric in the Einstein
frame is not affected by the duality twist and the corresponding
geometric flux must always be unimodular.}.

\section{Conclusions and open problems}\label{sec:conc}

In this paper we studied in detail the class of ${\cal N}=4$
axionic-symmetry gaugings and established that they can be
embedded in heterotic theory. More specifically, they arise
through a reduction where the boundary conditions for the fields
are twisted by an $SO(1,1)$ scaling symmetry. Similar reductions
are possible for type II strings  yielding ${\cal N}=8$ gauged
supergravities or ${\cal N}=4$ upon appropriate
orbifolding/orientifolding. For M-theory, instead, there is no
scaling symmetry of the action and that implies that the
Lagrangian cannot be consistently truncated for fields with
boundary conditions of this type. However, one can still perform
such reductions at the level of the equations of motion and it is
expected that the reduced equations of motion correspond to
gaugings of the type we studied here. In passing, we also note
that since the dilaton becomes a component of the internal
geometry when a type IIA background is lifted to M-theory,
reductions with dilaton twists should lift to M-theory reductions
with purely geometric twists of the type studied in
\cite{Dall'Agata:2005fm, Hull:2006tp}.

Some obvious extensions of the current work include twisted
reductions of the heterotic theory taking into account the
ten-dimensional gauge fields or similar type II reductions in the
presence of branes and orientifolds. This should yield electric
${\cal N}=4$ gaugings where the gauge algebra is a subgroup of
$SL(2,\mathbb{R})\times SO(6,n)$ for $n\geq 6$.

The fact that the formulation of gauged supergravity through the
embedding tensor is duality-covariant implies that these theories
capture the effective dynamics of backgrounds related by duality
transformations. Recently it has become increasingly  clear that the majority of
these backgrounds are non-geometric and cannot be described using
the familiar notions of geometry and ordinary fluxes. From one
point of view this demonstrates the power of the effective
bottom-up approach, since four-dimensional physics can be derived
without the need to delve into the microscopic details of a
higher-dimensional setup. On the other hand, one could argue that
a better understanding of non-geometric backgrounds may still be
obtained through analyzing gauged supergravity.

For instance, the ``non-geometric" fluxes $Q$ and $R$ proposed in
\cite{Shelton:2005cf} as T-dual of the familiar NS-NS and
geometric fluxes, are automatically captured for heterotic
compactifications on a six-torus by the formulation of ${\cal
N}=4$ gauged supergravity we have being discussing. Using the
notation of  \cite{Shelton:2005cf} our gauging parameters
$f_{+IJK}$ describe all possible situations through
\begin{equation}
f_{+ijk} \sim H_{ijk},\quad f_{+ijk'} \sim
f^{\hphantom{ij}k}_{ij},\quad f_{+i' j' k} \sim
Q^{ij}_{\hphantom{ij}k}, \quad f_{+i' j' k'} \sim R^{ijk}.
\end{equation}

Besides this set of $SO(6,6)$-dual fluxes,
the most general ${\cal N}=4$ gauging comprises of another set
of S-dual fluxes $f_{-IJK}$. It would be extremely interesting to understand
the microscopic origin of all those non-geometric fluxes directly
in ten dimensions and derive the corresponding gauged supergravities
using an appropriate reduction scheme (see \cite{Dabholkar:2005ve}
for some recent ideas in this direction. Also,
the non-geometric fluxes can be interpreted
as geometric fluxes in an appropriate generalized geometry \cite{dpst}).
Among others, this should shed some light on the open
problem of lifting gauged supergravities with non-trivial duality phases
in heterotic string theory.

A related question concerns the higher-dimensional origin of the
gauging constraints (\romannumeral1)-(\romannumeral4). For
example,  although some of these constraints have a clear origin
as Bianchi identities in the internal space, this is not so for
the null condition (\romannumeral1) $\xi_M \xi^M=0$. The reduction
we performed depends naturally on six parameters that fill up
$\xi_M$ in such a way that it is automatically null. It would be
interesting to understand how ${\cal N}=4$ gaugings with more
general parameters $\xi_M$ can be obtained from higher dimensions
and where the null condition comes from.

We conclude by emphasizing that formulations of string and
M-theory of the type presented in \cite{Hull:2004in, Hull:2006va,
Hull:2007zu} as well as the mathematical framework of
generalized complex geometry
\cite{Gualtieri:2003dx} may provide the appropriate tools for
resolving the above issues.


\bigskip

\begin{acknowledgments}
  The authors wish to thank L.~Carlevaro, G.~Dall'Agata, S.~Ferrara and H.~Samtleben
  for stimulating scientific discussions and J.~Schön, M.~Trigiante and M.~Weidner
  for very helpful correspondence.
  This work was
  supported in part by the EU under the contracts MEXT-CT-2003-509661,
  MRTN-CT-2004-005104, MRTN-CT-2004-503369 and by the Agence Nationale pour la
  Recherche, France, contract 05-BLAN-0079-01.
  Marios Petropoulos acknowledges financial
  support by the Swiss National Science Foundation.
\end{acknowledgments}

\appendix

\end{document}